\documentclass[fleqn,usenatbib]{mnras}

\usepackage[T1]{fontenc}

\usepackage[T1]{fontenc}
\usepackage{natbib}
\usepackage{graphicx}
\usepackage{amsmath}
\usepackage{amssymb}
\RequirePackage{fix-cm}

\usepackage{newtxtext,newtxmath}

\title[Dynamically Produced Moving Groups]{Dynamically Produced Moving Groups in Interacting Simulations}

\author[P. Craig et al.]{
Peter Craig,$^{1}$
Sukanya Chakrabarti,$^{1,2}$
Heidi Newberg,$^{3}$
Alice Quillen$^{4}$\\
$^{1}$School of Physics and Astronomy, Rochester Institute of Technology, 84 Lomb Memorial Drive, Rochester, NY 14623, USA\\
$^{2}$Institute of Advanced Study, Princeton NJ 08540, USA\\
$^{3}$Department of Physics, Applied Physics and Astronomy, Rensselaer Polytechnic Institute, Troy NY 12180, USA\\
$^{4}$Department of Physics and Astronomy, University of Rochester, Rochester NY 14627, USA}

\date{Accepted XXX. Received YYY; in original form ZZZ}

\pubyear{2021}

\begin{document}
\label{firstpage}
\pagerange{\pageref{firstpage}--\pageref{lastpage}}
\maketitle

\begin{abstract}
We show that Smoothed Particle Hydrodynamics (SPH) simulations of dwarf galaxies interacting with a Milky Way-like disk produce moving groups in the simulated stellar disk. We analyze three different simulations: one that includes dwarf galaxies that mimic the Large Magellanic Cloud, Small Magellanic Cloud and the Sagittarius dwarf spheroidal; another with a dwarf galaxy that orbits nearly in the plane of the Milky Way disk; and a null case that does not include a dwarf galaxy interaction. We present a new algorithm to find large moving groups in the $V_R, V_\phi$ plane in an automated fashion that allows us to compare velocity sub-structure in different simulations, at different locations, and at different times.  We find that there are significantly more moving groups formed in the interacting simulations than in the isolated simulation.  A number of dwarf galaxies are known to orbit the Milky Way, with at least one known to have had a close pericenter approach.  Our analysis of simulations here indicates that dwarf galaxies like those orbiting our Galaxy produce large moving groups in the disk.  Our analysis also suggests that some of the moving groups in the Milky Way may have formed due to dynamical interactions with perturbing dwarf satellites. The groups identified in the simulations by our algorithm have similar properties to those found in the Milky Way, including similar fractions of the total stellar population included in the groups, as well as similar average velocities and velocity dispersions.

\end{abstract}

\begin{keywords} 
    galaxies: dwarf -- galaxies: structure -- galaxies: interaction -- galaxy: kinematics and dynamics -- galaxy: solar neighborhood
\end{keywords}

\section{Introduction}

Moving groups have been observed in the local region of the Milky Way for many years. Many of these groups are quite well known, including the Hyades, Pleiades, Coma Bernices and Sirius moving groups \citep{Eggen}. Parallaxes and proper motions from the Hipparcos satellite \citep{hipp} were particularly useful for identifying disk moving groups \citep{Dehnen}.  Radial velocity surveys, in addition to the proper motions from Hipparcos, provide full 3D velocity information for stars in the Solar neighborhood \citep{hipp_data}, which is useful in the search for moving groups, where some detection methods require the full 3D velocity information. Finally, with the advent of Gaia data, our ability to search for moving groups near the Sun has increased significantly \citep{2018A&A...616A...1G}. Here we use a subset of the Gaia DR2 RVS sample, which includes all of the stars with low parallax errors that satisfy $\varpi / \sigma_\varpi > 5$, where $\varpi$ is the parallax and $\sigma_\varpi$ is the uncertainty on the parallax, and that are within 0.5 kpc of the Galactic plane.

A number of different methods have been used to identify moving groups.  A commonly used method for identifying moving groups in the Milky Way (MW) is the convergent point method. The idea is that the proper motions for stars that are in a group that is extended on the sky and that have similar 3D space velocities will converge towards a single point on the sky; the groups will be extended on the sky if they are either large or nearby. Looking for these convergent points can enable the identification of many structures in the MW. This method is biased towards including stars that have small proper motions, which is not ideal for our data. Many of our data sets include a large number of particles that would have small proper motions because they are relatively far away from the Sun, which could lead to many false positives in our results \citep{Eggen1965,deBruije1999}. Our method would suffer the same problem if used on a set of stars with proper motions only, however it is more readily applied to the 3d space velocities, which are always available in our simulations. Another method is to use a wavelet decomposition on the velocities of the stars in the sample, which also has been shown to detect many of the moving groups in the Gaia DR2 RVS sample \citep{2016A&A...595A...1G, 2018A&A...616A...1G, 2018A&A...616A...7S, 2019A&A...622A.205K,wavelet}.  As we discuss below, our method identifies moving groups by fitting Gaussians to regions in velocity space that have a high density of stars, relative to the background stellar velocity distribution.

Several mechanisms have been suggested to explain the formation of moving groups. A common explanation is that these velocity structures are the remnants of open clusters, or formed by interactions with a bar \citep{Dehnen_Bar,Eggen1965}. One problem with the cluster formation idea is that stars in moving groups can have a variety of different ages and compositions, so it is unlikely that they all came from the same cluster \citep{Eggen1965,Kushniruk2020}.  Analysis of GALAH data \citep{2018MNRAS.478..228Q} indicates that some moving groups, such as the Hercules moving group, may be due to a resonant bar.  It has also been suggested that moving groups could have been formed from perturbations due to the Magellanic Clouds via gravitational interactions \citep{Dehnen}.  Recent work also finds that transient spiral structure \citep{2018MNRAS.481.3794H} may lead to the formation of moving groups, as well as perturbations due to spiral arms in the MW \citep{Michtchenko}. Moving groups in Gaia data have also been identified and analyzed in action space. In the ($J_R$ , $J_z$) plane there are at least seven overdensities that follow lines of constant slope in this plane, which correspond to known moving groups in the solar neighborhood \citep{Trick}. It is likely that there may be multiple causal mechanisms at play in the formation of moving groups in the Milky Way. The analysis of Gaia DR2 data has revealed many facets of a Galaxy that are clearly out of equilibrium, including the so-called phase-space spiral \citep{antoja2018}, and the Enceladus merger \citep{helmi2018}, that have been interpreted as arising from interactions with dwarf galaxies.  Analysis of Gaia DR2 data also led to the discovery of a new dwarf galaxy \citep{torrealba2019} that likely interacted with the Milky Way \citep{chakrabarti2019}.  However, the formation of moving groups due to dwarf galaxy interactions has not yet been studied with full \emph{N-body} simulations.  Motivated by these earlier works that indicate that the MW may has been perturbed by dwarf galaxies, we focus our study here in trying to understand if some of the moving groups in the Galaxy may have arisen from dwarf galaxy interactions.

Here, we explore the formation of moving groups in Smoothed Particle Hydrodynamics (SPH) simulations of the Milky Way interacting with satellites, and contrast these cases with an isolated simulation to study the differences in their evolution and structural appearance. By carrying out full N-body simulations of Milky Way like galaxies that include stars, gas, and a live dark matter halo, some of which are perturbed by dwarf galaxies, we go beyond prior work that has used equilibrium models to analyze moving groups, isolated simulations, or test particle calculations. Earlier work \citep{2009MNRAS.397.1599Q} used test particle calculations to study the formation of moving groups in an interacting simulation with a perturbing satellite.  Hybrid techniques (incorporating both massive and tracer particles) used to simulate the Galactic disk also produce moving groups that resemble the Hercules stream \citep{2011MNRAS.417..762Q}. It is important to emphasize that in contrast to works such as \cite{wavelet} that have identified a comprehensive list of moving groups in Gaia DR2, we are not attempting to present an exhaustive list of all the moving groups, but rather are focusing on identifying the large and prominent moving groups.

This paper is organized as follows.  In \S 2, we review the simulation methodology, and in \S 3 we review the algorithm that we use here to characterize moving groups. In \S 4, we apply our moving group algorithm to Hipparcos and Gaia DR2 data and show that it does recover a number of large known moving groups. In \S 5, we discuss the general characteristics of the moving groups found in our simulations, contrasting the interacting and isolated simulations. In \S 6, we qualitatively compare moving groups formed in these simulations to those found in the Milky Way.  We discuss and present our conclusions in \S 7.

\section{Simulations}

We analyze three Milky Way-like simulations, two that include dwarf galaxy interactions and one without a satellite.  We performed these simulations with the parallel TreeSPH code GADGET-2 \citep{2005MNRAS.364.1105S}. GADGET uses an N-body method to simulate the collisionless components (the dark matter and stars) and Smoothed Particle Hydrodynamics (SPH) to follow the evolution of the gaseous component, including the formation of new stars. The gravitational softening lengths are 100 pc for the gas and stars and 200 pc for the dark matter halo particles.  In the primary galaxy, we use $4 \times 10^{5}$ gas particles, $4 \times 10^{5}$ stellar particles and $1.2 \times 10^{6}$ dark matter halo particles. The model is initialized with a stellar mass of $3.5 \times 10^{10} M_{\odot}$, a gas mass of $1.25 \times 10^{10} M_{\odot}$, and a dark matter mass of $1.44 \times 10^{12} M_{\odot}$.  These Galactic components are represented in the simulation by different particle types: gas, disk, halo (dark matter) and new star particles (that are created during the simulation from gas particles according to the Kennicutt-Schmidt prescription \citep{2003MNRAS.339..289S}. Disk particles represent the stellar component of the galaxy which is present at the start of the simulation; these stars do not evolve over time. Our version of GADGET-2 uses an effective equation of state \citep{2003MNRAS.339..289S} where a sub-grid model is adopted for energy injection from supernovae. We adopt a value of 0.75 for the bulk artificial viscosity parameter and the details of the implementation of the viscosity are described in \cite{2005MNRAS.364.1105S}.

As in prior work \citep{Chakrabarti&Blitz}, we use a Hernquist profile matched to the NFW profile \citep{2005MNRAS.361..776S} to simulate the dark matter halo of the Milky Way-like galaxy.  We use a concentration of 9.4, a spin parameter of $\lambda = 0.036$, and a circular velocity $V_{200} = 160$ km s$^{-1}$.  We also include an exponential disk of gas and stars, as well as an extended HI disk.  The disk mass fraction is 4.6\% of the total mass, which results in a radial scale length for the exponential disk of 4.1 kpc.  The mass fraction of the extended HI disk relative to the total gas mass is 0.3, with a scale length that is three times that of the exponential disk \cite{Chakrabarti&Blitz}. The satellites are set up using similar methods, and also use a Hernquist profile matched to the NFW profile.

Simulations that include dwarf galaxies have observationally motivated orbits for the dwarf galaxies. Starting with the observed positions and velocities of observed Milky Way dwarf galaxies at present day, we carry out a backward integration in a Galactic potential (the Hernquist profile matched to NFW as described in \cite{2005MNRAS.361..776S}) using an orbit integration code \citep{2011MNRAS.416..618C} for several Gyr, until the desired starting time is reached. The dwarf galaxies are comprised of dark matter, disk, and gas particles in the beginning of the simulations, and can form new star particles during the simulation. The co-planar dwarf is comprised only of dark matter particles, so there are no gas or disk particles and thus no star formation in this satellite. The LMC and SMC are set up with all three components, and include star formation. The Sagittarius dwarf includes dark matter and disk particles, but does not contain any gas. We evolve the simulation forwards with these initial conditions for timescales of several Gyr. In this way we create a Milky Way-like simulation in which the simulated dwarf galaxies are close to the present day positions of their counterparts at a known time after the beginning of the simulation. The co-planar simulation is evolved forward in time for a total of $2$ Gyr, the three dwarf galaxy simulation for $5.8$ Gyr, and the null simulation for $3$ Gyr. We analyze the simulation output files at intervals of $50$ Myr.

In one simulation, we simulate the Sagittarius (Sgr) dwarf galaxy, the LMC and the SMC. We will refer to this simulation as the Sagittarius-like simulation because it is believed that the Sgr dwarf has the largest influence on disk substructure. The initial conditions and orbits of the three satellites have been derived using HST proper motions \citep{2017AAS...22912307L}. The progenitor mass of the Sgr dwarf in this simulation is $2.89 \times 10^{10} M_{\odot}$, and the progenitor masses for the LMC and SMC are $1.90 \times 10^{10} M_{\odot}$ and $0.19 \times 10^{10} M_{\odot}$, respectively. In this three dwarf simulation, the simulated Sgr dwarf experiences two pericenters over the last three Gyr, with the closest pericenter to present day occurring 0.3 Gyr after present day, where it reaches a distance of 13 kpc. The LMC and SMC do not approach closer than $\sim$ 50 kpc relative to the Galactic disk over the last few Gyr, as found also in earlier work \citep{2012MNRAS.421.2109B}. We have listed our derived initial conditions (the starting positions and velocities) for the satellites in Table 1. These positions and velocities are for the satellites at the beginning of the simulation, and are the results of our backwards orbit integration.

\begin{table*}
    \centering
    \begin{tabular}{| c | c | c | c | c | c | c | c | c | c |}
    \hline\hline
     Object & x (kpc) & y (kpc) & z (kpc) & vx (km s$^{-1}$) & vy (km s$^{-1}$) & vz (km s$^{-1}$) & Start Time (Gyr) & \shortstack{Pericenter \\ Time (Gyr)} & \shortstack{Pericenter \\ Distance (kpc)}\\
     \hline
    Sagittarius & 29.6 & 0.58 & -15.8 & -66.6 & 30.1 & 164.6 & -5.8 &0.3 & 13\\
    LMC & 48.8 & 259.2 & -67.0 & -31.7 & -235.4 & -2.80 & -5.8 & -0.1 & 57\\
    SMC & -10.7 & 184.2 & 4.6 & 40.3 & -136.4 & 81.7 & -5.8 & -.25 & 48\\
    Co-planar & 97.5 & 21.2 & 1.2 & -251.8 & -23.8 & 0.2 & -2 & -0.3 & 8\\
    \hline
    \end{tabular}
\caption{This table includes the initial positions and velocities for the satellites in our simulations, as well as the start times, pericenter times and distances. All times listed are relative to present day, wit negative times occurring before present day and positive times after present day.}
\end{table*}

The co-planar simulation that we analyze here is the same one that is described in \cite{Chakrabarti&Blitz}, i.e., it includes a massive dwarf ($M_{\rm dwarf} \sim 10^{10} M_{\odot}$) progenitor with an orbital plane that is nearly in the disk \citep{Chakrabarti&Blitz}, and a close pericenter approach ($\sim$ 8 kpc). This close pericenter approach occurs $0.3$ Gyr before present day. The parameters of the co-planar simulation performed by \cite{Chakrabarti&Blitz} are similar to (but are not exactly the same as) those done recently by \cite{2019arXiv190604203C}, which describe a dynamical model for the newly discovered Antlia2 dwarf galaxy.

The third simulation does not have a perturber, and we use this ``null" simulation as a point of contrast to the interacting galaxy simulations. The simulated MW galaxy for the null case is produced with the same initial conditions as the primary galaxy in the simulations that include dwarf galaxies.

\begin{figure}
\begin{center}

\includegraphics[width=\columnwidth]{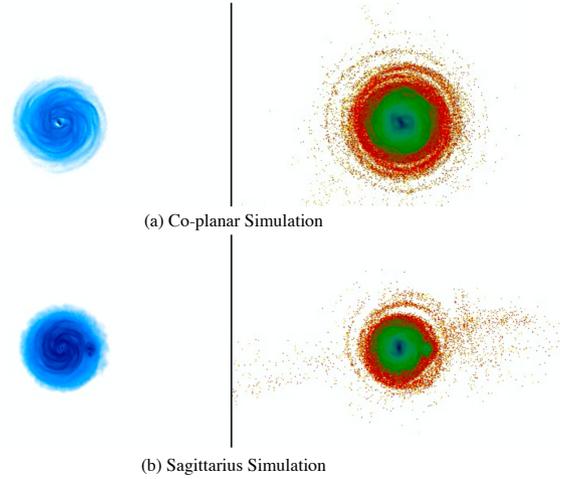}

\caption{a. (Left) Gas density map at present day in the co-planar simulation.  (Right) Stellar density map at present day in the co-planar simulation.
b. (Left) Gas density map at present day in the Sagittarius-like simulation.  (Right) Stellar density map at present day.}\label{f:3sats}
\end{center}
\end{figure}

Face-on renderings of the simulated Milky Way at present day for the gas and stellar density distribution are shown in Figure~\ref{f:3sats} for both the co-planar simulation and for the Sagittarius-like simulation. These figures show the formation of spiral structure in a stable disk. The Milky Way in the null simulation also includes similar spiral structure inside of a stable disk. The spiral and ring like features seen in Figure \ref{f:3sats} also appear in the null simulation. Figure~\ref{f:3sats} shows the tidally stripped Sgr dwarf galaxy superimposed on the disk. Note that the other satellites are not visible here as they are far from the center of the MW at the present time. The Fourier amplitudes for the spiral structure in the co-planar simulation can be found in \cite{Chakrabarti&Blitz}. Typical values for the Fourier amplitudes of the gas surface density range of the low-order modes in the outer HI disk are $\sim$ 0.2, with comparable values for the stellar disk.  It is likely that over the time-scales explored in this paper, where moving group formation correlates closely with the interaction with the satellite, the role of gas dynamics (i.e., dissipation and angular momentum transport to the stars) is not significant.

\section{Algorithm} \label{sec:floats}

We developed a method to find and characterize groups in various data sets in an automated fashion, given the large quantity of data and the need to have objective measures for comparison. This method is designed to identify the largest velocity structures in a data set within a reasonable computation time. It is not designed to detect every group that exists, and in fact is likely to miss the smaller structures that have been identified in the Solar neighborhood. Methods that have been applied to finding groups in the Solar neighborhood are reviewed in \S 1. Our method will, however, reliably detect large moving groups, which is what we require for this work. Studying smaller structures is difficult without running very high resolution simulations.  Given that our goal is to carry out a global analysis of velocity structures in SPH simulations to see if prominent moving groups are formed by dwarf galaxy interactions, we do not require a more finely-tuned algorithm for our purposes. We have applied our algorithm to observed stars with reliable three dimensional velocities and positions in the Solar neighborhood, so that the identified groups can be compared with groups found in the stellar components of each simulation, selected from regions of each simulation that are similar to the Solar neighborhood in distance from the Galactic center. We will show that our algorithm identifies the prominent groups in real data for the Solar neighborhood that were found in previous works. For each identified group in both the real data and the simulations, the average velocity, velocity dispersion, and stellar fractions are determined.  

We fit a series of Gaussian distributions to the density of stellar bodies as a function of $(U,V)$ velocity, where $U$ points towards the center of the Milky Way, $V$ points in the direction of motion of a typical star on a circular orbit, and $W$ points out of the plane in the direction of the positive $Z$ axis. U and V are measured with respect to a local Cartesian coordinate system moving with the LSR. It is worth noting that the algorithm will also work if provided with Galactocentric cylindrical coordinates instead of U and V, which is a more appropriate choice for some applications. $U$, $V$ and $W$ are replaced by $U = -V_R$, $V = -V_\phi - V_{circ,\odot}$, and $W = V_z$, where $V_R, V_\phi,$ and $V_z$ are right-handed Galactocentric cylindrical coordinates.  Throughout this work we have used the values for the solar motion given in \citep{Schonrich2012}. At the solar radius, $V_{circ,\odot} \approx 236 $ km s$^{-1}$, as given in \citep{vcircsource}. The velocity data is binned into a 2D histogram in $U$ and $V$, with pixels of 1 km s$^{-1}$ per side. Note that the pixel sizes used in the figures are generally not the same, as different pixel sizes may make better figures.

We first fit one large, axis-aligned Gaussian, with a center and width determined by the average and standard deviation of the data set in each direction. This first Gaussian represents the velocity dispersion of the disk. Then, this best fit Gaussian profile is subtracted from the number of stars in each bin so that we can look for local overdensities that we identify as ``moving groups".  

We search the residual for regions in the $U-V$ plane that have an appreciably higher density than the best fit single Gaussian profile. We then fit additional axis-aligned Gaussian distributions, usually containing fewer stars than the original Gaussian, and with smaller standard deviations. Each of these Gaussian distributions describes one of the moving groups within the data set. It is worth noting that in principle, any vertex deviation is possible and allowed for the moving groups identified. It is difficult to accurately measure this with the number of particles found in a typical group. The effects of any vertex deviation are not expected to make a significant impact on the algorithm performance in our simulations. Each of the Gaussian distributions here has a total of five parameters: a center location in the U and V directions, standard deviations in these directions, and one parameter for the amplitude of the Gaussian. The center parameters are simply set by finding the center of an over-density. In this case the center of the overdensity is assumed to be located at the pixel with the highest number of particles above the background. Any pixels adjacent to this with a positive value after background subtraction will be considered to be part of the overdensity, as well as positive pixels adjacent to the adjacent pixels. This information is used only for the purposes of estimating the standard deviations. If this yields only 1 or two pixels, we will reduce the bin size in this area to estimate the standard deviations, as the correct standard deviation may be smaller than the pixel size. The number of particles contained in the over density can then be used to appropriately set the amplitude of this Gaussian.

Next the algorithm uses an optimization routine to improve the accuracy of these parameters. The optimization uses a 2d Kolmogorov-Smirnov (KS) test implemented in python to evaluate the goodness of fit \citep{Peacock1983,Fasano1987} of our distribution. Most of the parameters are reasonably close to the correct values already, except for the initial standard deviations, which is remedied by the optimization routine. The optimization uses scipy minimization functions to minimize the value for the KS test of this group. The chosen minimization routine uses a gradient descent algorithm. These optimizations are done for the groups one at a time, where the parameters for all of the other distributions are held constant. Ideally, we would be able to fit them all at once, but this is computationally expensive. An exception occurs if two group centers are close enough that they are likely to be fit by overlapping Gaussians such that we need to optimize them simultaneously; in this case the parameters of both Gaussians are fit simultaneously. This condition is not common, but can be triggered if the two groups have estimated centers that are within 10 km/s of each other.

At this point, the algorithm will have identified the locations of every large group in the data set, but it also will have false positives that need to be removed. We do this by comparing the number of stars in the central region of the group to the number we expect in that region from the background. If the number of stars in this region is significantly larger than what is expected to randomly occur, then we will classify the structure as a group. The error in the background is estimated by the square root of the number of particles in that pixel. The residual is the number of particles in the pixel, in excess of the background counts.  A residual value more than twice the error means that without the presence of the new group, the pixel deviates significantly from the background model. The $2\sigma $ requirement is only relevant in the higher density region of a moving group, where the overdensity will be the most significant. In the velocity wings of the background Gaussian distribution, where the Poisson error in the background is small (often a single star), spurious groups are identified by the requirement of a minimum of three stars (or particles) in a group. The requirement of a minimum of three particles was selected to eliminate any false positives at large distances from the background, where a single particle may be significantly above what is expected from the background, and anything smaller than three particles may not represent a real moving group. If any pixel in a given group satisfies the $2\sigma $ criterion \emph{and} the minimum object count requirement, then we accept the group as real.  Thus, even though the criterion is applied to every pixel, not every pixel in an extended structure must satisfy this requirement.  Typically, in groups spanning a wide range in $U$ and $V$, only one of the central pixels of the group satisfies our criteria; pixels farther out in the velocity tails of the group do not, and do not have to, satisfy any of these criteria. These criteria were tested on the task of identifying known structures in the Hipparcos data set, and were found to effectively eliminate many false positives, while still identifying known structures that were identified by other authors with other techniques. The three star minimum introduces a weak dependence on the number of stars, which could make the comparison between the Milky Way and the simulations difficult, as the number of particles in the simulations within the solar neighborhood is much less than the number of stars in the solar neighborhood with data in Gaia DR2.  However, as we discuss below, we find in practice that this is only a problem for very low contrast moving groups.  

If the background distribution is significantly different from our background fit, then we will have problems accurately determining the significance of the groups. In this case, one region of the simulation will have a higher sensitivity than it would otherwise, while other regions will have a lower sensitivity. The importance of this effect changes based on how many stars are in the sample; with large numbers of stars the effect is small relative to the scale of the groups detected. This can be a consideration when the true background is not well modelled by a Gaussian distribution.

A potential concern with a method like ours arises from asymmetric drift. It is well known that asymmetric drift significantly modifies the local velocity distribution \citep{2013A&A...557A..92G}. Due to the negative gradient in density and velocity dispersion, there are more stars with smaller guiding radii near apocenter in the solar neighborhood than there are stars with larger guiding radii near pericenter. We use a K-S test to show that the impact of asymmetric drift is small in our simulated data.  Specifically, we check that in regions of the simulation that do not contain any large velocity structures (i.e. that are reflective of the background), a two sided K-S test yields p-values that typically range from 0.2 to 0.65.  Thus, we cannot reject the hypothesis that the background is symmetric. Asymmetric drift will shift the mean of the distribution and add a skewness in the tangential component. The shifting of the mean will not impact the algorithm, given that the mean for the background will be determined based on the data. The skewness is potentially problematic for this algorithm, as it will make the distribution deviate from the assumed Gaussian background distribution. This effect has been tested to be irrelevant for the larger groups in the dataset, where the effect of the lower quality background fit is small relative to the size of the groups. Asymmetric drift will be important in a search for small groups with this method, where the number of particles in the group is comparable to the error in the background distribution. Large groups will be far less likely to be impacted, as the size of these structures will be significantly larger than any error in the background. For the groups we are interested in, the effects of asymmetric drift will likely be small relative to the size of the over density. The effect may be present, but it will not hinder our ability to detect large scale velocity sub-structures.

A justification for the decision to focus on large moving groups is as follows. Many of the identified groups in the MW have very low mass relative to the resolution in our simulations. For example, many of the groups listed in \citep{wavelet} contain under 1000 stars. The total mass of these groups is likely less than that of a single particle in our simulations, and as such would be undetectable. Thus it is not useful for our purposes to identify structure on these scales or smaller in the MW. One way to decrease the density contrast of moving groups that could be detected is by using a larger survey volume, which would include more stars in each group and thus increase the total mass and the detectability within the simulations. However, this will only be effective if the moving group is coherent throughout the volume.

\begin{figure}
\begin{center}
\includegraphics[width=\columnwidth]{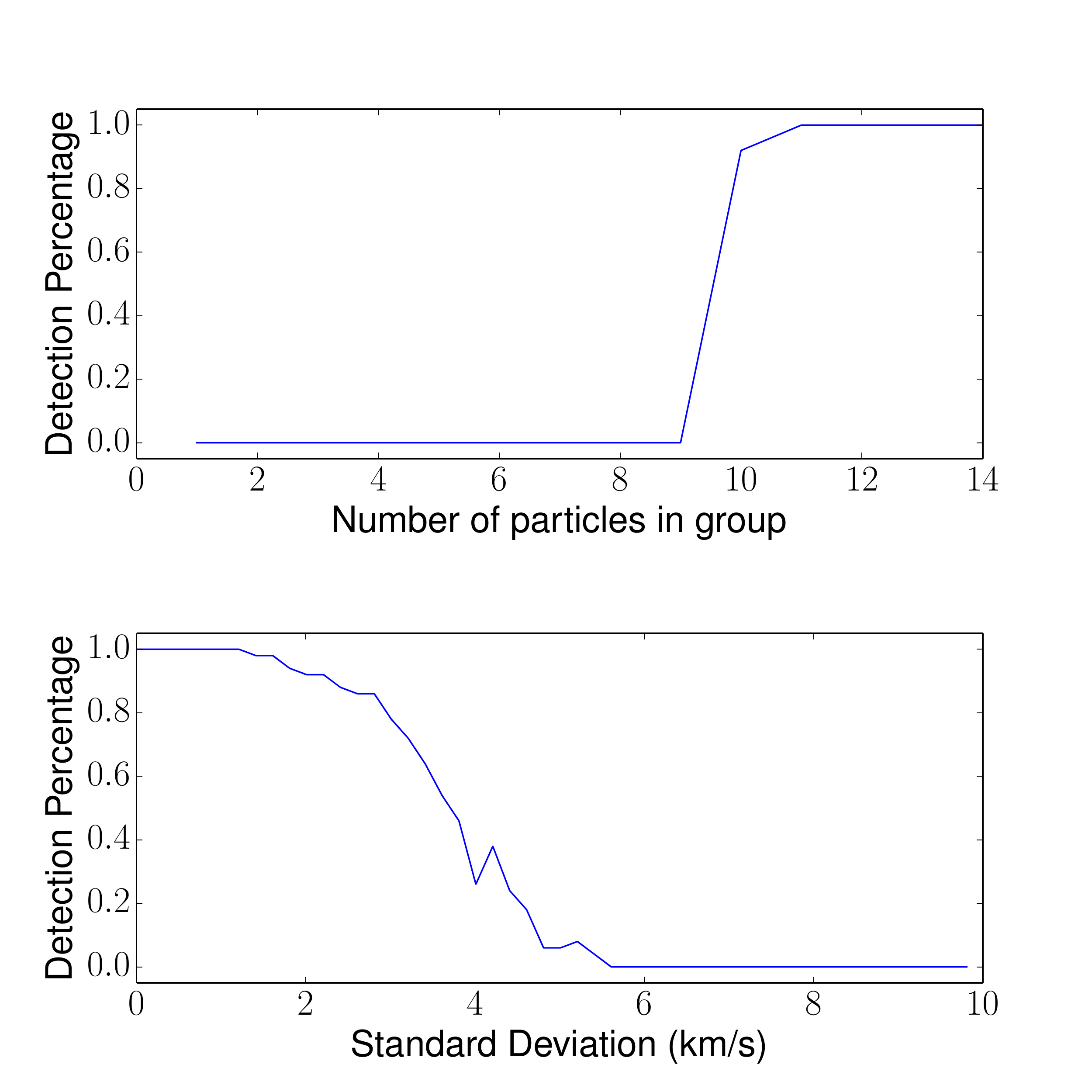}\\
\caption{This figure shows the performance of the algorithm as the number of particles and standard deviation of a group is changed. In each case we generate 50 random groups with identical parameters, and then determine what percentage of these groups are successfully recovered by the algorithm. In both cases we place a single group into a background distribution taken from the initial snapshot from the co-planar simulation, located at a distance from the center of the background of 25 km s$^{-1}$. The dispersion of the background in U is 25.39 km s$^{-1}$ and the dispersion of the background in V is 21.90 km s$^{-1}$.}\label{f:perf}
\end{center}
\end{figure}

\begin{figure*}
\begin{center}
\includegraphics[width=\textwidth]{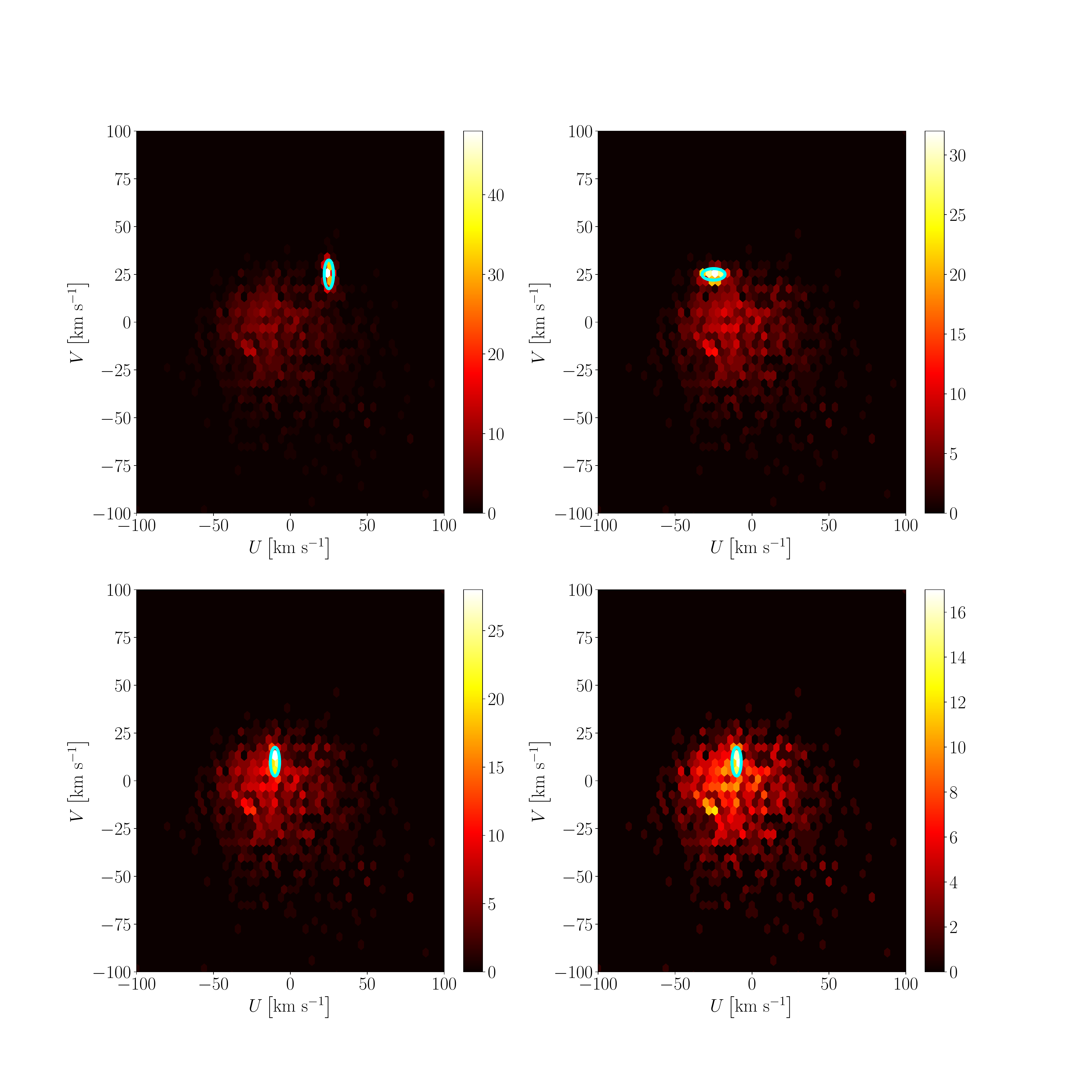}\\
\caption{This figure shows 4 sample groups placed on top of a background taken from the beginning of the co-planar simulation. In each case, the cyan ellipses show the position and standard deviation of the detected groups. In this figure the size of the ellipse is set to three times the standard deviation, so that it contains most or all of the particles in the group. The top left panel shows a very large group, comprised of 400 particles, with a large standard deviation. The top right panel contains a group with half the number of stars, and the standard deviations and position changed. In the bottom left panel we reduce the number of particles to 100 and move the test group closer to the center of the background. In the bottom right panel, we include a group containing 45 particles. The white ellipses show the location of the background distribution, with the heights and widths of the ellipses set as twice the standard deviation in that direction.}\label{testing}
\end{center}
\end{figure*}

We have carried out numerous tests to evaluate the performance of the algorithm. The tests are carried out by artificially introducing groups that are reflective of what we expect the simulations to produce, on top of a background taken from an early epoch of the simulation with a nearly co-planar dwarf galaxy. A basic test that we perform is to test whether a single group with a hundred stars and a small standard deviation can be identified in a background stellar distribution with a thousand stars. Such a group will always be correctly identified.  Additional tests that we have carried out include placing groups in different places relative to the background with different numbers of stars and different standard deviations. The general tendency is that the algorithm produces better results when groups have large numbers of stars, small standard deviations, and are far from the center of the background. The threshold for recovering moving groups depends on three things: the number of particles in the group, the standard deviation, and its placement relative to the center of background. The dependence on the number of particles and the standard deviation can be seen in Figure \ref{f:perf}. The dependence on the distance from the background is not shown as this depends heavily on the shape of the background distribution, and this factor is typically less significant than the other two. Thus, while we can reliably recover a group of ten particles (or even fewer) with a small standard deviation far from the background, we would not be able to do so if such a group were placed close to the center of the background.  Far away from the center of the background, small structures containing three or more stars could potentially be identified as moving groups, however such structures are uncommon.  We find that most of the identified groups contain a minimum of 10 simulation particles, which corresponds to our detectability threshold. The results of several tests are shown in Figure \ref{testing}. This figure shows the velocity distribution for test data containing large velocity structures on top of background data from the simulations. In each case an ellipse marks the location of every structure identified by the algorithm, showing both the center location and the standard deviation of the identified group. The algorithm recovers each group that was placed in these test data sets with parameters that are close to the correct values.

\section{Group Finding in the Milky Way}

In order to confirm that this algorithm is effective in finding moving groups in real data, we have used it to analyze observations of the Milky Way. We first looked for groups in the Hipparcos data. Here, our goal was to confirm that the algorithm is able to identify groups within the subset of the Hipparcos data that has 3D velocities compiled from other catalogues, and parallax errors that are less than $ 10 \%$.  We then check whether the parameters for the groups agree with those that have been previously obtained. We used the Pulkovo compilation \citep{hipp_data} of radial velocities for Hipparcos data. This data set can be used to reproduce groups that have been studied in the Milky Way, and provides an initial sample set for testing the algorithm. It is worth noting that some of the stars in this sample have a large error in their positions and velocities. This is reduced by the parallax error cut, which filters out many of the stars that have highly uncertain positions, which are often a large factor in the 3D velocity uncertainties. 

\begin{figure}
\begin{center}
\includegraphics[width=\columnwidth]{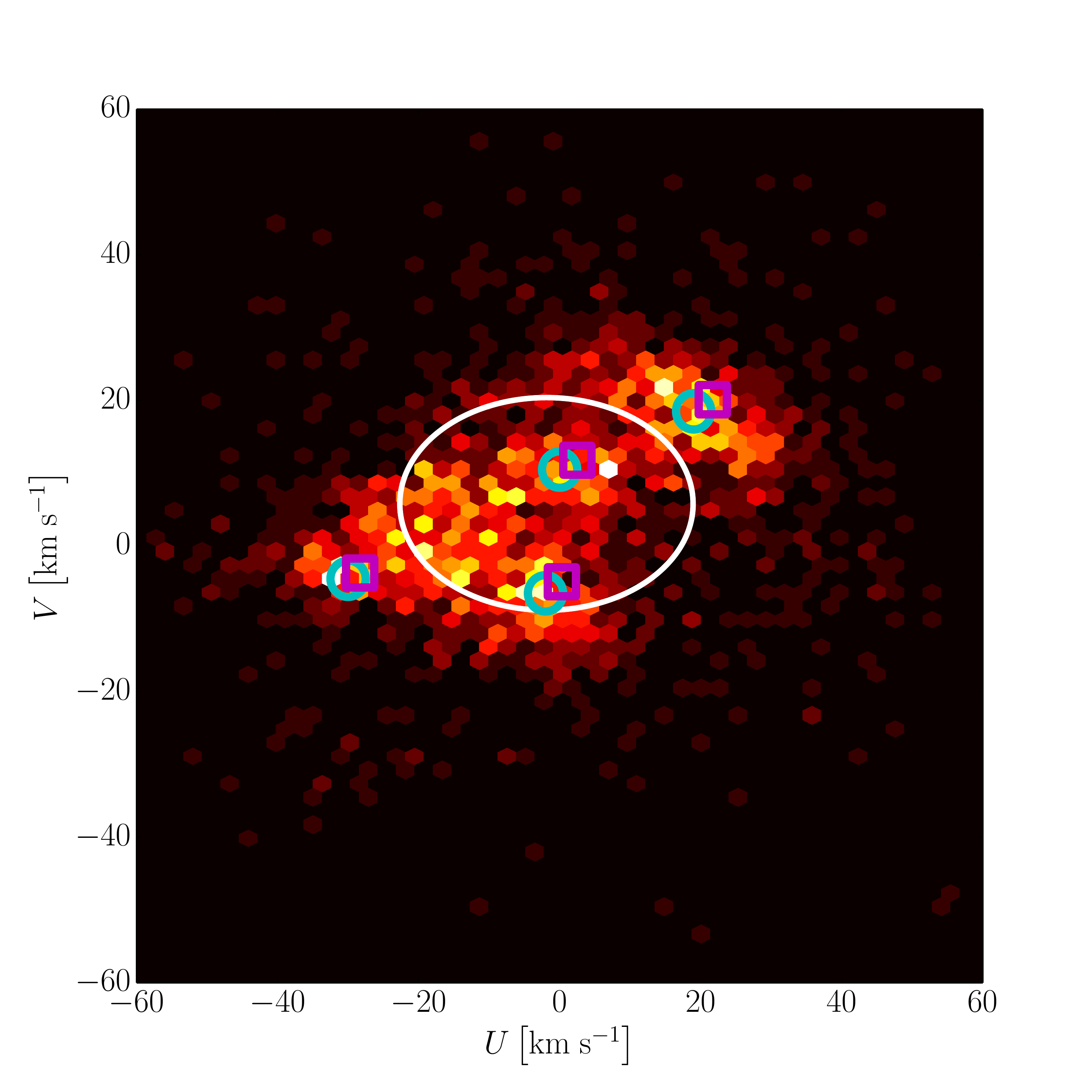}
\\
\caption{Velocities of stars in the Hipparcos catalogue. Here we show the positions of known moving groups in the Milky Way \citep{Dehnen} and compare them to the groups detected by the algorithm. The cyan circles show the positions of known groups from \citet{Dehnen}, while the magenta squares show the locations of the same groups output by the algorithm. The white ellipse shows the location of the background distribution, with the height and width of the ellipse set as twice the standard deviation in that direction.}\label{fig:hipp}
\end{center}
\end{figure}

Figure~\ref{fig:hipp} shows the $U,V$ velocity distribution for Hipparcos stars, with a sample size of 35,489 stars. In this plot there are four groups that are easily identified, and several rather small groups that are not detected by our algorithm.  We have compared the $U, V$ values associated with these groups with earlier work by \citet{Dehnen}, and we find close agreement. The results of our algorithm when run on this data set are shown in Table \ref{pulkovo_table}, along with the group parameters listed in \cite{Dehnen}. 

\begin{figure}
\begin{center}
\includegraphics[width=\columnwidth]{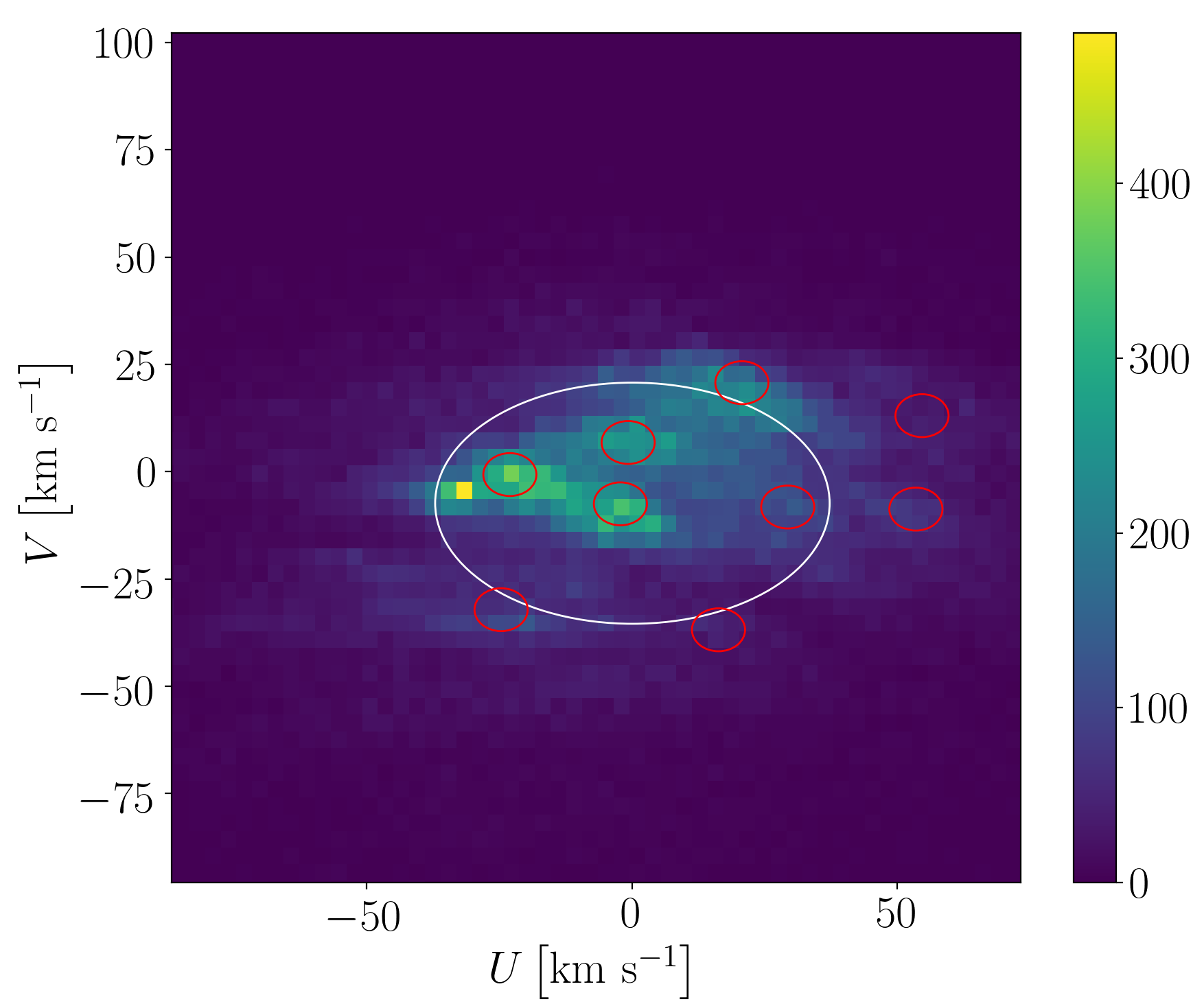}\\
\caption{Velocities of stars in the Gaia DR2 RVS sample. The red circles mark the locations of the 9 groups detected by our algorithm in this data set. The bright yellow pixel is likely either a star cluster or an extension of the Hyades moving group. We have not been able to identify a particular cluster with that velocity, but it remains a likely explanation. The white ellipse shows the location of the background distribution, with the height and width of the ellipse set as twice the standard deviation in that direction.}\label{Gaia}
\end{center}
\end{figure}

We also tested our algorithm against the same Gaia DR2 sample that was used in \citep{wavelet}. This sample is generated by starting with the entire Gaia DR2 RVS sample. We then select sources with low parallax uncertainties, where $\varpi / \sigma_\varpi > 5$. We also limit the sample to stars with Galactocentric Z positions between 0.5 kpc and -0.5 kpc. This leads to a sample containing 5,136,533 stars with known velocities. For some of our testing we use this full data set, while for others we take a region with a radius of 100 pc. The second test cases are the most useful as they allow a direct comparison to the groups discussed in \cite{wavelet}, which also makes this cut when identifying groups in the Solar neighborhood. In the simulations, there are many spherical volumes at the Sun's distance from the Galactic center that can be compared with real data in the solar neighborhood. Our algorithm detects nine statistically significant groups in the \cite{wavelet} data, that are listed in Table \ref{t1}, along with the central velocities of our groups and the corresponding groups found in \cite{wavelet}. A figure showing the U and V distribution of this data set can be seen in Figure \ref{Gaia}, along with the recovered positions of the moving groups displayed in red circles. This is only a small sample of the list of groups detected in \citep{wavelet}, which identifies 44 groups in the MW. It is worth noting that many of the groups identified in this paper are too small to be detected in our simulations. This table shows that when we carry out our analysis on the RVS sample from Gaia DR2, we do reproduce the \cite{wavelet} locations to within reasonable tolerances (most are within 1 km s$^{-1}$) for the nine large moving groups that our algorithm detects. The $\gamma$leo moving group, which is smaller than the others listed in Table 2, shows a larger deviation from the \cite{wavelet} values, as might be expected with the application of our algorithm, which was designed to work on large moving groups.  

\begin{table*}
\centering
\begin{tabular}{| c | c | c | c | c | c | c | c |}
\hline\hline
Name & $U$ (km s$^{-1}$) & $V$ (km s$^{-1}$) & Algorithm $U$ (km s$^{-1}$) & Algorithm $V$ (km s$^{-1}$) & $\sigma_U$ (km s$^{-1}$) & $\sigma_V$ (km s$^{-1}$) & Fraction\\
\hline
Pleiades & -2 & -6.5 & -0.9 & -5.9 & 4.45 & 4.95 & 0.043\\
Hyades & -30 & -4.5 & -29.2 & -3.54 & 4.79 & 4.31 & 0.045\\
Sirius & 19 & 18.5 & 19.35 & 19.21 & 3.55 & 3.62 & 0.041\\
Coma Berenices & 0 & 10.5  & 0.2 & 10.64 & 4.56 & 4.85 & 0.034\\
\hline
\end{tabular}
\caption{This table shows the positions of four prominent moving groups identified in the Pulkovo compilation. The columns labelled U and V show the positions identified in the Hipparcos data in \citep{Dehnen}, while the algorithm U and V columns show our identified locations. The $\sigma_U$ and $\sigma_V$ columns provide the standard deviations for the groups measured by the algorithm. The fraction column displays the total fraction of the stars in the sample determined to be in the group.}\label{pulkovo_table}
\end{table*}

\begin{table*}
\centering
\begin{tabular}{| c | c | c | c | c | c | c | c |}
\hline\hline
Name & $V_R$ (km s$^{-1}$) & $V_\phi$ (km s$^{-1}$) & Algorithm $V_R$ (km s$^{-1}$) & Algorithm $V_\phi$ (km s$^{-1}$) & $\sigma_U$ (km s$^{-1}$) & $\sigma_V$ (km s$^{-1}$) & Fraction\\
\hline
Pleiades & 2.0 & -228.5 & 2.16 & -228.50 & 4.51 & 5.10 & .041\\
Hyades & 23.0 & -235.5 & 22.97 & -235.32 &  4.78 & 4.23 & .040\\
Sirius & -20 & -256.5 & -20.71 & -256.73 &  3.53 & 3.44 & .034 \\
Coma Berenices & -0.5 & -246.5  & 0.70 & -242.8 &  4.62 & 4.99 &  .046\\
Dehnen98-6 & -30.5 & -228.0 & -29.34 & -227.76 & 2.98 & 3.15 & .021\\
Dehnen98-14 & -54.0 & -227.0 & -53.47 & -227.28 &  2.64 & 2.71 & .015 \\
$\gamma$leo & -63.5 & -253.5 & -54.62 & -249.05 &  2.67 & 2.34 & .010 \\
Hercules 2 & 24.5 & -203.5 & 24.62 & -203.84 &  2.59 & 2.15 & .016 \\
Liang17-9 & -16.5 & -199 & -16.31 & -199.15 &  2.24 & 1.89 & 0.008\\
\end{tabular}
\caption{This table shows the positions of moving groups determined by the algorithm within the RV sample of Gaia DR2. The names listed are obtained from \citep{wavelet}. $V_R$ and $V_\phi$ are previously determined velocities, while Algorithm $V_R$ and $V_\phi$ are determined by our algorithm. To make this comparison, we ran our algorithm on the same data set used in \citep{wavelet}. Note that in this case we compute the group parameters in cylindrical coordinates. It is expected that the choice of coordinates will not make a significant impact on the results of the algorithm.}\label{t1}
\end{table*}

\section{Groups in the simulations}

The fiducial region of the simulations that we sample in the following analysis is located at $X = 8$ kpc and $Y = 0$ kpc, and we select stars by choosing all stars within a sphere with a radius of 1 kpc from this location.  This selection is meant to simulate the region around the Sun in which we observe moving groups. With a disk resolution of $8 \times 10^{4} M_\odot$, the selection yields on the order of 1000 particles. The average number of particles in this region throughout the co-planar simulation is 1196. This is one of several regions sampled in each simulation, with similar behavior exhibited in axially symmetric places around the disk. All of the sampled regions are at a radius of 8 kpc from the Galactic center, and each of the sampled regions shows similar general trends. 

The stars included in these regions are only the stars that have existed since the beginning of the simulation, i.e., none of the newly generated stars are included. In this paper, we identify moving groups in our simulations only in this stellar population, i.e, for the disk stars.  We do not see significant differences in moving group formation if we include the new stars in addition to the old stars in our simulations. Using only newly formed stars to find moving groups is occasionally problematic, especially early in the simulations when there are not as many new star particles.  There are other times where the background distribution for the new stars is not well behaved; in particular there are cases where the Gaussian fit to the background is not a good fit to the actual background, especially earlier in the simulations when the number of newly created stars in the solar neighborhood is small. For this reason, we restrict our analysis to the old stars only.

\subsection{Groups at the Solar Circle}
\begin{figure*}
\begin{center}
\includegraphics[width=\textwidth]{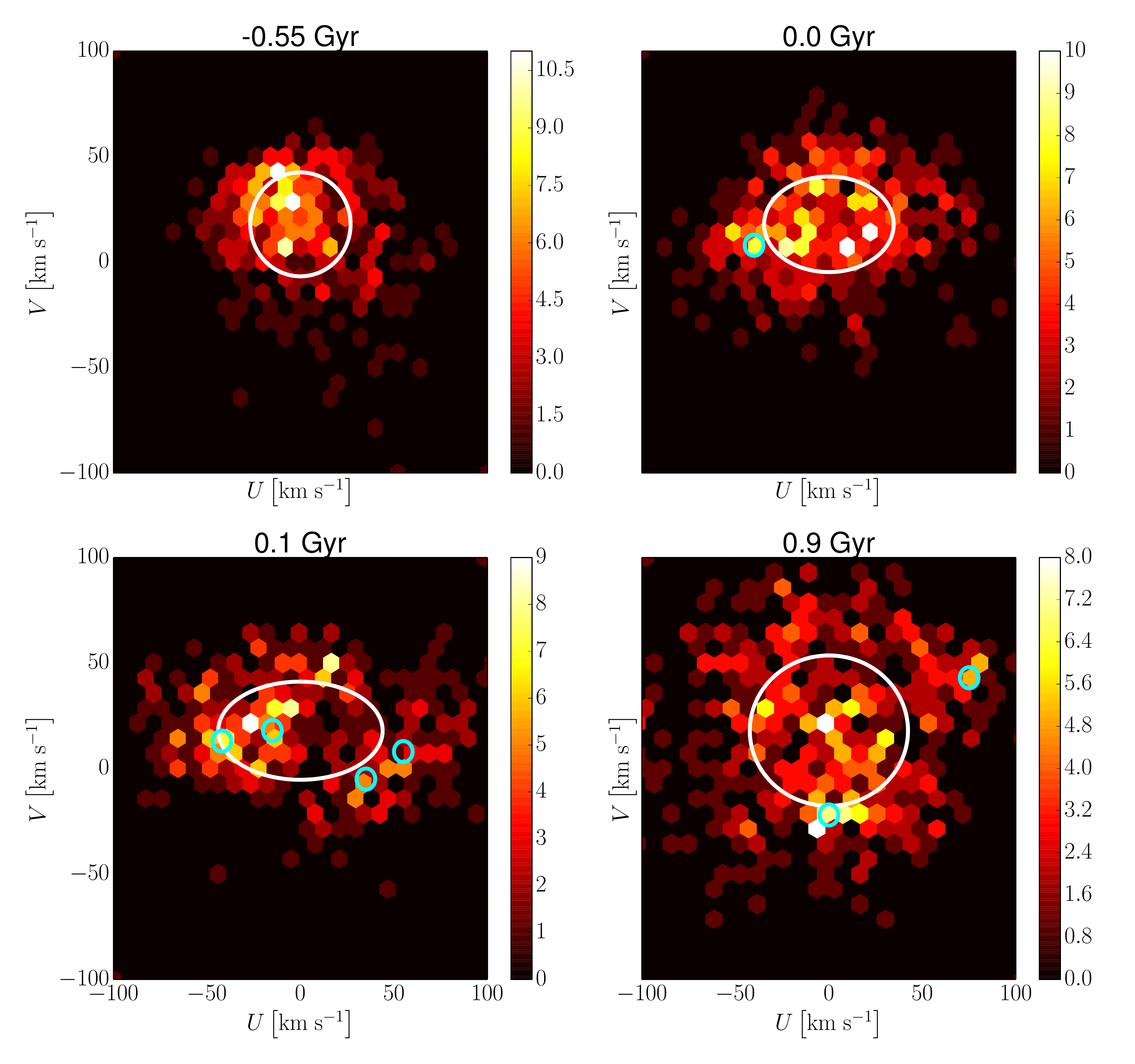}\\
\caption{$U$ and $V$ velocities for stars in the disk for the simulation with a co-planar dwarf galaxy.  The time labels are with respect to present day in the simulation. Here we see the clear formation of velocity substructure beginning near present day, which is shortly after the collision with the dwarf galaxy. The data shown is drawn from a spherical region of the disk 1 kpc in radius 8 kpc from the center of the galaxy. Similar results are obtained by using regions at different position angles within the disk. Note that the colors in this figure represent the total number of disk particles in a given sample in that velocity bin. The blue circles mark the positions of moving groups identified in this data set. The white ellipse shows the location of the background distribution, with the height and width of the ellipse set as twice the standard deviation in that direction.}\label{dr1}

\end{center}
\end{figure*}

\begin{figure*}
\begin{center}
\includegraphics[width=\textwidth]{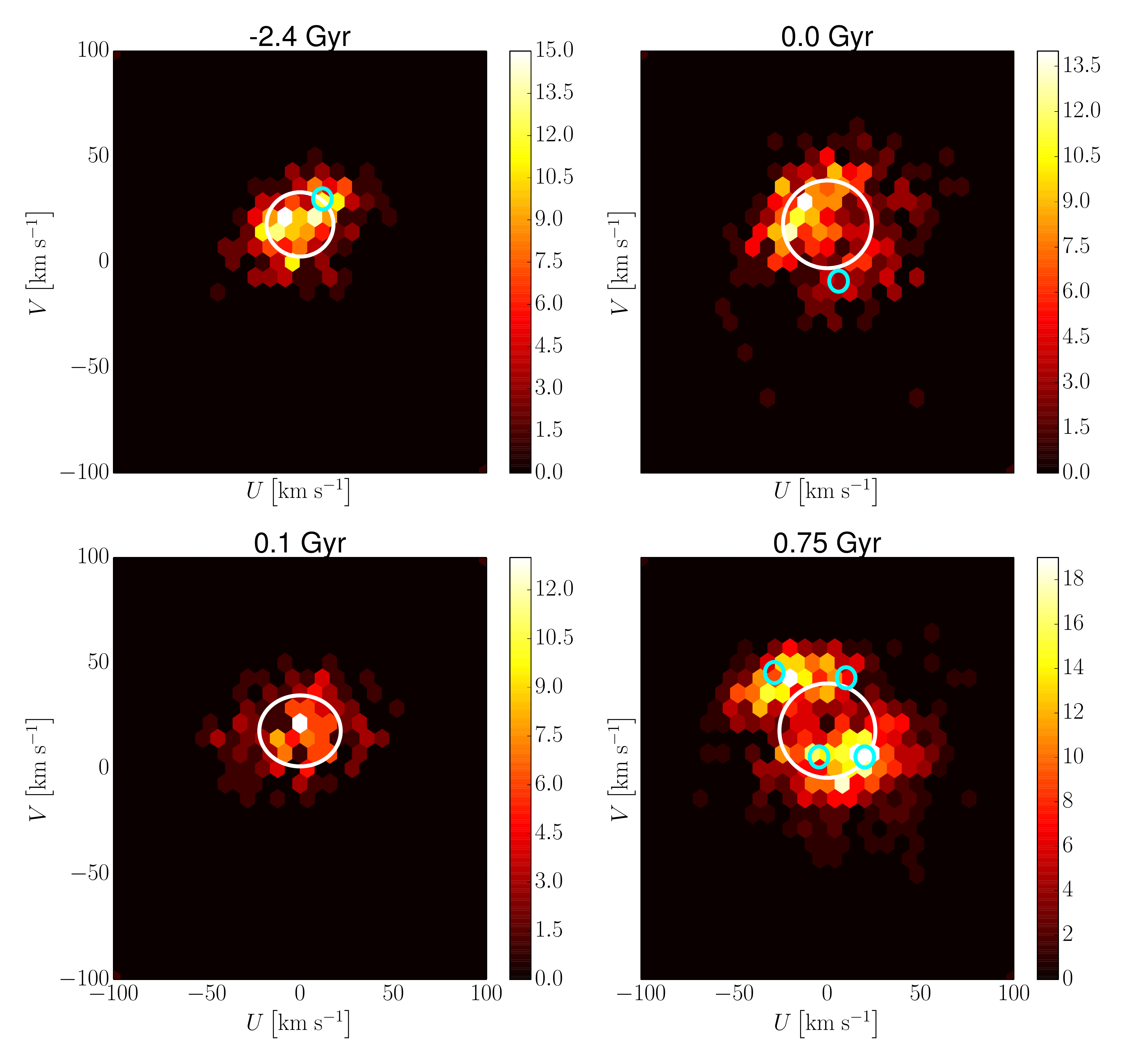}\\
\caption{This figure is similar to Figure \ref{dr1}, however using the simulation with three dwarf galaxies. The times shown are relative to present day, and we see similar results; there are clear velocity structures that form after present day in the simulations, which is again after the interaction has had time to perturb the disk. This is the same region of the disk as shown in Figure \ref{dr1}. Similar results can be obtained by gathering data from other angles in the disk in this simulation.}\label{sgr1}

\end{center}

\end{figure*}

\begin{figure*}
\begin{center}
\includegraphics[width=\textwidth]{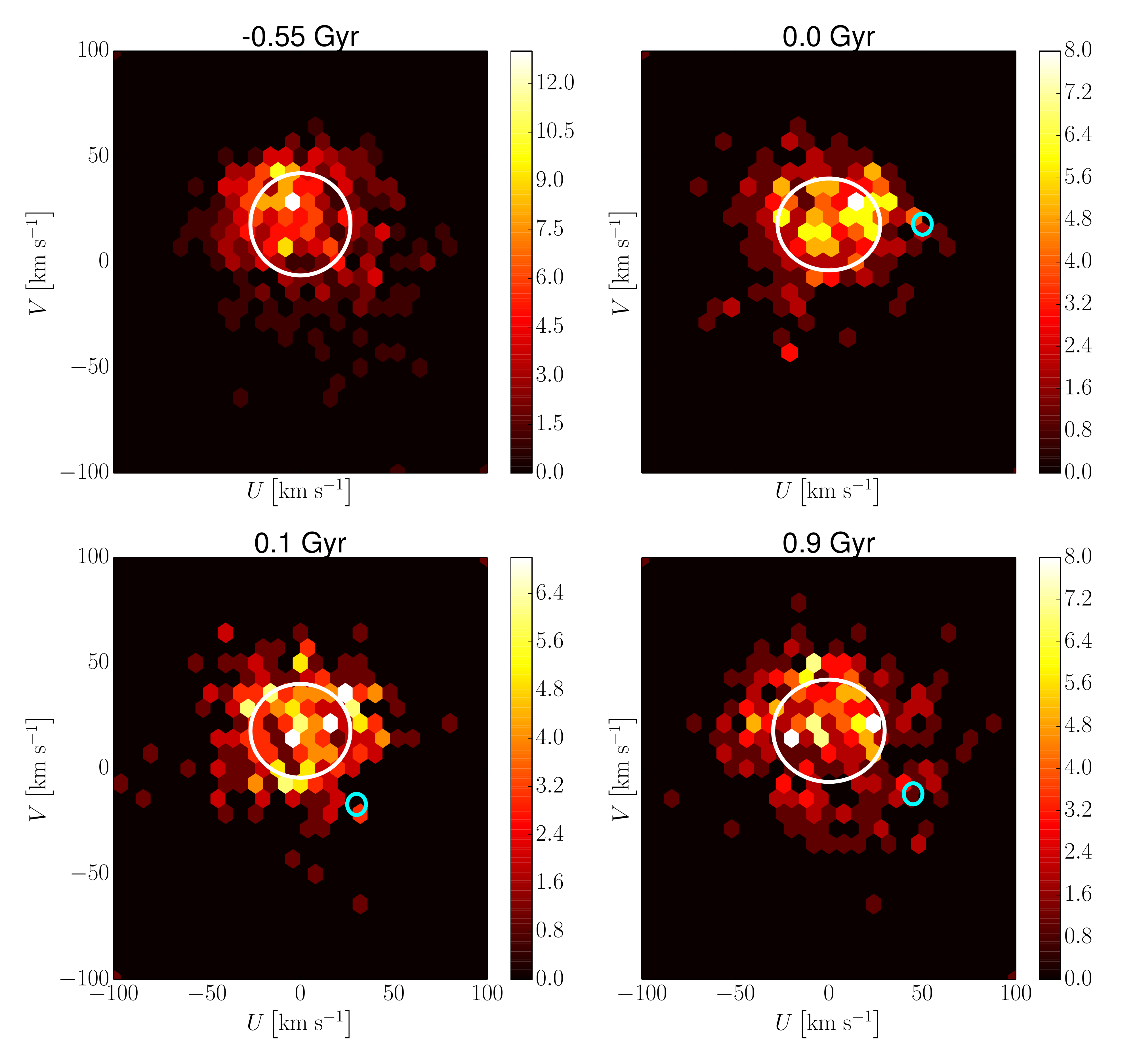}\\
\caption{This figure is similar to Figure \ref{dr1}, except that this time we are using the null simulation. Here we do not see the formation of many large groups, as we would expect. Similar plots made with times taken throughout the simulation yield similar results, which is a clear indication that the structures we see in our interacting simulations are a result of the perturbation from the dwarf galaxy. The region shown is the same as the region shown in the previous two figures.}\label{null1}

\end{center}
\end{figure*}

Figure \ref{dr1}, \ref{sgr1} and \ref{null1} show the distribution of $U$ and $V$ velocities for the disk stars in our co-planar, Sagittarius-like, and null simulations, respectively. The data displayed in these figures is sampled at our standard location described above. Each of the four panels shows the velocity sub-structure in that volume at a different time in the simulation. The panels are labeled with the time with respect to the present day, with negative times occurring before the present day in the simulations. Blue circles mark the central locations of moving groups detected in the simulations. Early in the simulations there has not been sufficient time for structures to form, so the earliest panels do not show any interesting velocity structures. At later times, we can clearly see interesting velocity structures form in both of the interacting simulations. There is much less prominent velocity structure formation in the null simulation. Below, we discuss these findings quantitatively. The lack of significant velocity structures in the null simulation while being present in the interacting simulations is consistent with the idea that interactions cause the formation of the moving groups that we see here.  


There is usually only limited group formation within the first 500 Myr of the simulation, before the perturber has made a close pass. The pre-interaction structure for the co-planar simulation can be seen in the early-time snapshots in Figure \ref{dr1}. The pericenter approach for the co-planar simulation corresponds to t= -0.3~\rm Gyr.  The dynamical effect is manifest roughly a dynamical time after this pericenter approach which is close to present day, i.e., $t \sim 0~\rm Gyr$, when there are $\sim$ four times as many moving groups as the maximum number identified in one region in the early-times. The simulation including three dwarf galaxies is shown in Figure \ref{sgr1}, which also demonstrates significant structure formation. Interestingly the amount of structure in this simulation increases after present day, which is closer to the second pericenter approach of the Sagittarius dwarf. The null case is shown in Figure \ref{null1}, which does not develop a large number of groups at any point in time.

There are typically variations with the azimuth angle selected in the disk at all Galactocentric radii sampled. However there is not a sufficiently clear trend to trace increased structure formation back to the position of the perturber in these simulations. In fact the variations seem to be random as the azimuth angle changes. It is most likely the case that in order to clearly see the variation of moving group structures as a function of azimuth, we would need to be able to identify significantly smaller structures, which is not possible with these simulations and the algorithm adopted here. It is possible that we would be able to see variations in the structures formed through the interaction with higher resolution simulations. Theoretically, we would expect to see some variation as a function of azimuth, since the azimuthally dependent effects of the perturber should be evident with both the larger and smaller groups. However, there are not enough large groups available to reach a statistically significant conclusion with the available data.

\subsection{Properties of the Groups}

We have compiled a number of statistics describing the groups found in these simulations. These statistics are determined from a compilation of 80 different data sets obtained from the simulations. We select spherical, 1 kpc regions centered at a location 8 kpc from the Galactic center with a $Z$ value of 0 kpc. In each simulation snapshot, we select data from 8 locations around the disk over several output files, from 100 Myr before present day to 100 Myr after present day, in both the co-planar and Sagittarius-like simulations. These groups have an average standard deviation of 1.33 km s$^{-1}$ in the $U$ direction and 1.24 km s$^{-1}$ in the $V$ direction. These are typically small compared to the mean $U$ and $V$ positions for the groups, which are almost entirely located at least 10 km s$^{-1}$  from $U$ and $V$ of 0.

In the null case there is an average of 0.1 moving groups per data set at the same Galactocentric radius as the Sun, which is significantly lower than what we see in the interacting simulations. The null case never generates more than one moving group, while both interacting simulations can produce several in the solar neighborhood, which indicates that the perturbations are causing the formation of additional velocity structures.

Groups identified in the simulations typically contain something on the order of $1 \%$ of the particles in the sample. The smallest groups detected contain just under 10 particles, which with typical simulation volumes on the order of 1000 particles yields a fraction of 0.01. In solar neighborhood samples, the fractions can range up to 5 $\%$ for a single group, depending on the number of stars in the sample. In a typical sample, between $2 \%$ and $10 \%$ of the simulation particles are contained within a moving group. At different radii the number of particles per group remains roughly constant, but the fractions change slightly because of variations in the sample size.

\begin{figure}
\begin{center}
\includegraphics[width=\columnwidth]{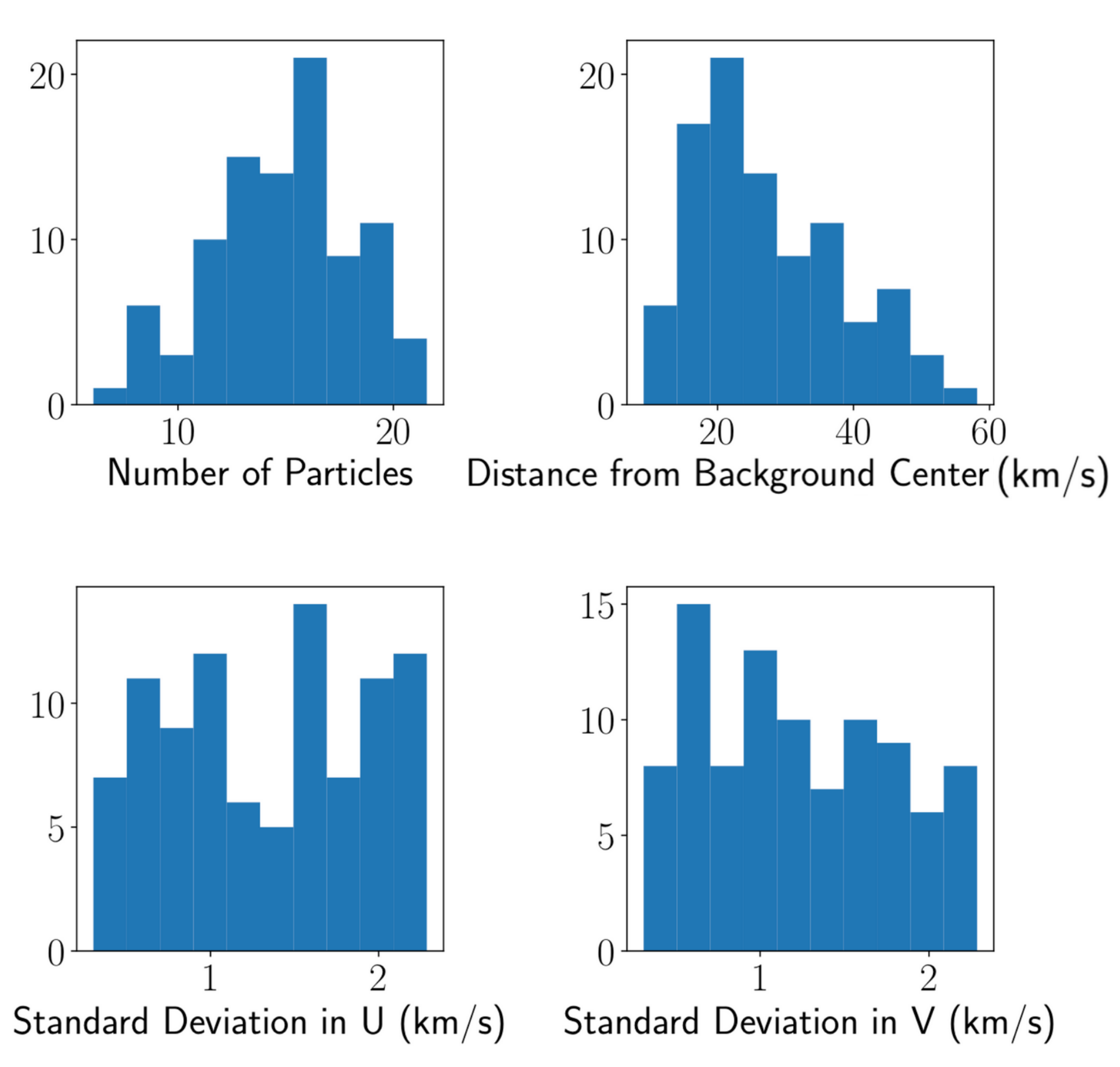}\\
\caption{Four histograms showing the distributions of assorted group properties, specifically the number of particles in the groups, the distance from the center of the background, and the standard deviation in each direction.}\label{histogram}
\end{center}
\end{figure}

Histograms showing the distributions of these quantities can be seen in Figure \ref{histogram}, which includes all the groups in the data sets described above. The number of particles identified in a given group is typically around 15, and ranges from 5 to 22. Typically the structures that we are finding are within 20 km s $^{-1}$ of the center of the background distribution, with the number of structures found steadily falling as the distance increases. This is to be expected as there typically are fewer particles at larger distances. The standard deviations are spread relatively evenly, but this is in part an effect of the algorithm's detection efficiency which is better for structures with small standard deviations as they present higher contrast overdensities.

\subsection{Groups as a Function of Radius}

\begin{figure}
\centering
\includegraphics[width=\columnwidth]{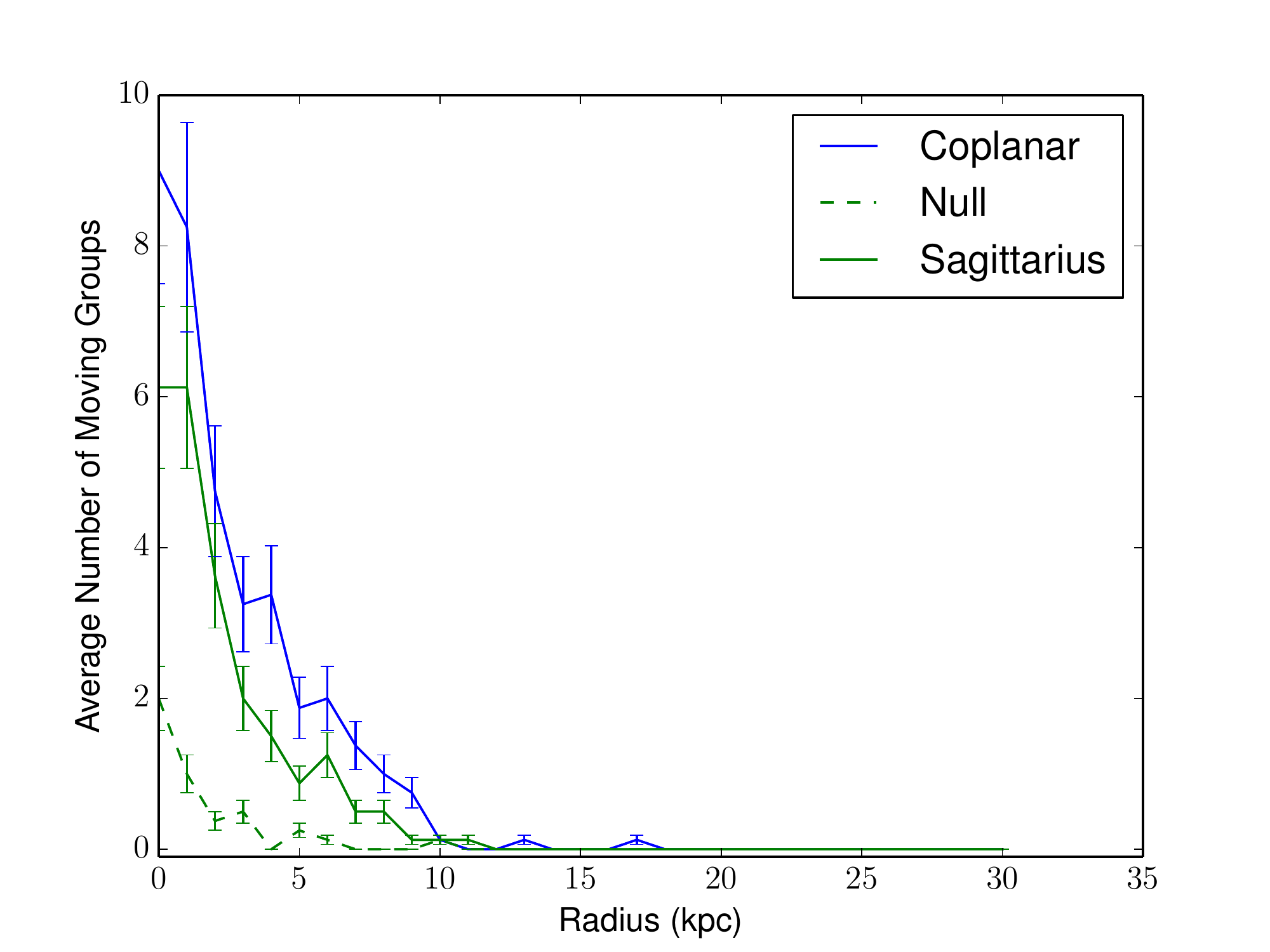}\\
\caption{Average number of groups as a function of radius at present day. The results for the null simulation are plotted at the present day for the Sagittarius simulation. This shows that the two interacting simulations contain more groups detected by the algorithm than the null simulation does throughout the disk, indicating that many of the detected groups are a result of the perturbation. The error bars show the Poisson errors on each point, which indicate that the null simulation and the interacting simulation do not agree within error.} \label{radius}
\end{figure}

It is interesting to examine the number of groups as a function of Galactocentric radius. If the moving groups are caused by dwarf galaxy interactions, then the null simulation is expected to have a smaller number of substructures across all radii relative to the interacting cases. Figure \ref{radius} depicts the average number of groups, as a function of distance from the galaxy center, in each simulation at the present day. The number of groups displayed is an average over several different locations, selected to have the same Galactocentric radii but at different azimuths. Since Figure~\ref{radius} shows the average number of groups over multiple different azimuths, the values shown in this figure at a Galactocentric radius of 8 kpc are not exactly the same as the number of groups shown in Figures \ref{dr1} , \ref{sgr1} and \ref{null1}. In the inner regions of the disk, we switch from $U$ and $V$ to $V_R$ and $V_\phi$.  In these areas we can no longer operate in a local Cartesian system, especially at $R \sim 0~\rm kpc$ where there is no good way to select a circular velocity; switching to a local Cartesian coordinate system is not viable when there are particles with very different azimuths within the chosen volume.  It is worth noting that the differences between these coordinate systems will not impact the algorithm, as the shapes of the distributions in velocity space do not change under this coordinate transformation.   
 
We calculate the average number of moving groups shown in Figure \ref{radius} by selecting several different angles in the disk, then measuring the number of groups at a given radius across all of the angles and averaging those values. Farther than 2.5 kpc from the galaxy center, we use eight equally spaced azimuthal positions. At radii less than 2.5 kpc, we reduce the number of azimuths to 4 in order to avoid an overlap between the spherical volumes. Note that the number of groups at each location in the disk is variable, so at the higher radii where the average number of groups is low, there are still some locations containing a larger number of groups.  Each individual measurement is made using all of the disk stars located within a sphere with a radius of 1 kpc.  As is clear, the interacting simulations have a larger number of moving groups relative to the null simulation, \emph{and} a larger number of moving groups at all radii out to 10 kpc from the Galactic center. The error bars on Figure \ref{radius} show the Poisson errors for each point, calculated using the number of groups detected. From this it is clear that the number of groups identified in the interacting simulations is significantly larger than the number identified in the null simulation. It is worth noting that we can produce significantly more structure in the interior regions of the disk. For instance there are typically about 8 groups at a radius of 2 kpc for the co-planar simulation, while at 8 kpc we never recover more than 4. The number of groups identified at radii larger than 10 kpc is quite small. This is possibly due to the smaller number of particles in samples taken at these radii, which makes it more difficult to recover moving groups.

\begin{figure}
\begin{center}
\centering
\includegraphics[width=\columnwidth]{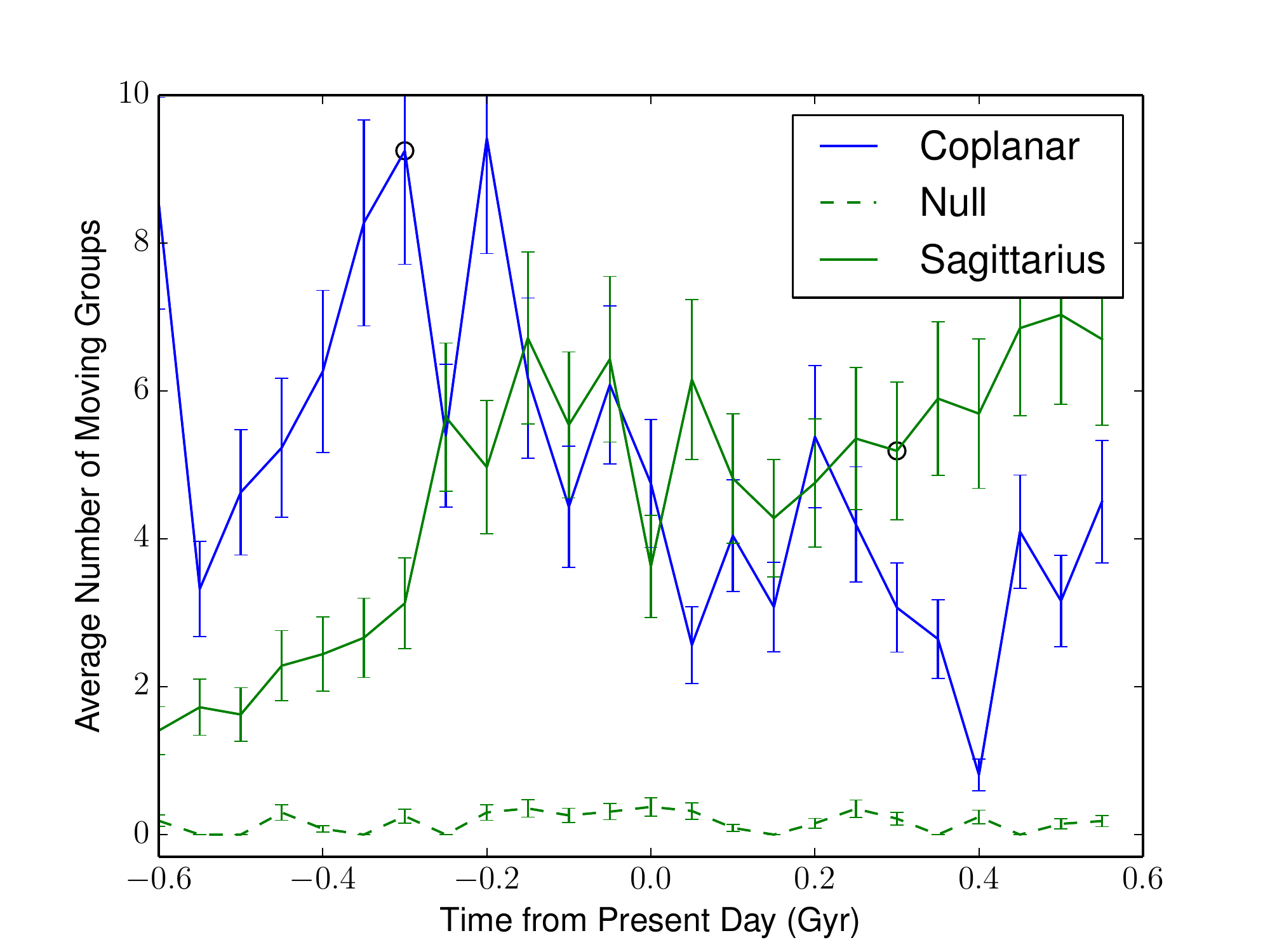}\\
\caption{Average number of groups at 2 kpc as a function of time centered at present day. Here the two green curves (representing the Saggitarius and null simulations) are plotted at the present day of the Sagittarius simulation and the blue curve is at the present day of the co-planar simulation. The time chosen for the null case is arbitrarily selected, and the time selected does not significantly alter the results. There is some time dependence here in both interacting simulations, but no time dependence in the null case. The circles mark pericenter passages closest to present day in each simulation. The error bars show the Poisson errors on each point, which indicate that the null simulation and the interacting simulation do not agree within error.}\label{time}
\end{center}
\end{figure}

\subsection{Groups as a Function of Time}
In a similar fashion, we can look at the time dependence of the average number of groups, which is shown in Figure \ref{time}. The times for the pericenter approach for each simulation are marked with a circle.  Note that we do not identify a particular time corresponding to present day for the null case, so for this case we have chosen the time of present day to be the same as in the Sagittarius simulation; the results do not change significantly with other time selections. Here the error bars represent the Poisson errors on each data point, which show that the number of groups in the interacting simulations is significantly larger than what is found in the null simulation. We can not track an individual moving group over time, as the groups form and dissipate on timescales less than 50 Myr, which is the time between our simulation snapshots.

Here we use the same averaging procedure used for examining the radial dependence, with the number of groups calculated at four different locations around the disk at a Galactocentric radius of 2 kpc. This region is currently inaccessible for observational moving group searches in the Milky Way. It was chosen for our simulation analysis because it has more large velocity structures than regions that are farther out in the disk. The number of observed structures in the simulations that correspond to regions of the Galactic disk that are currently accessible with observed data is small, which makes a search and analysis for time dependent trends difficult.

The number of groups in the Sgr and co-planar simulations is comparable to a null simulation at early times ($t \sim -1$ Gyr or earlier), and increases appreciably after the interaction. This is what one would expect if the groups are caused by the dwarf galaxy, as it takes about a dynamical time for the gravitational perturbation to make a significant impact on the disk. On this scale, the point of closest approach in the Sagittarius simulation occurs 0.3 Gyr after present day. There is another pericenter passage from the Sagittarius dwarf galaxy 1.05 Gyr before present day, which is close enough to cause significant perturbations. In the CB09 (co-planar) simulation, the pericenter occurs 0.3 Gyr before present day.

\section{Comparison of Simulations with the Milky Way}

There are a large number of moving groups that are well known in the Milky Way that can readily be compared to the moving groups in our simulations. Many such groups were originally found in Hipparcos data, and more recently can be seen in Gaia data. Such groups include the Pleiades, Hyades, and Coma Bernices. In tests of our algorithm, these groups were easily detected in both the Hipparcos and Gaia data for stars near the Sun. There are some groups that prove to be problematic for the algorithm to identify, such as the NGC 1901 group, which is fairly small and close to the center of the background distribution. This group represents a weak point of the algorithm, which is that the algorithm has greater difficulty in identifying groups near the center of the background distribution. It is still generally successful in these regions, but occasionally struggles on small groups like NGC 1901. In the region of the disk that corresponds to the Solar neighborhood, we see fewer large structures than what is observed in the MW, while we observe larger amounts of structure in the interior regions of the disk.

Comparing the groups found in simulations with these groups in the Milky Way is not straightforward because the number of stars in the Solar neighborhood in the Milky Way is larger than in the comparable region in the simulations. One difference is that we do not see many very small groups in the simulations, but there are many in the MW, which is likely a combination of our resolution and the algorithm's inability to accurately detect smaller structures \citep{Dehnen}. We can compare the fraction of stars or particles present in the largest groups detected between samples. In principle this should be readily comparable between our simulations and the MW. We recover similar group fractions between our results and the MW for the groups detected, but recover a lower total fraction, likely because there are many more detected groups in the MW. The MW contains a number of groups containing small fractions, which are not detectable by our algorithm. The groups that we do detect in the MW contain $5 \%$ or less of the total sample, with detected groups containing 23.1 $\%$ of the stars in the Gaia Solar Neighborhood sample. This is slightly underestimated as there are many smaller known groups which are not included, although the impact of these groups is small as they do not contain a large number of stars.

\begin{figure}
\begin{center}
\includegraphics[width=\columnwidth]{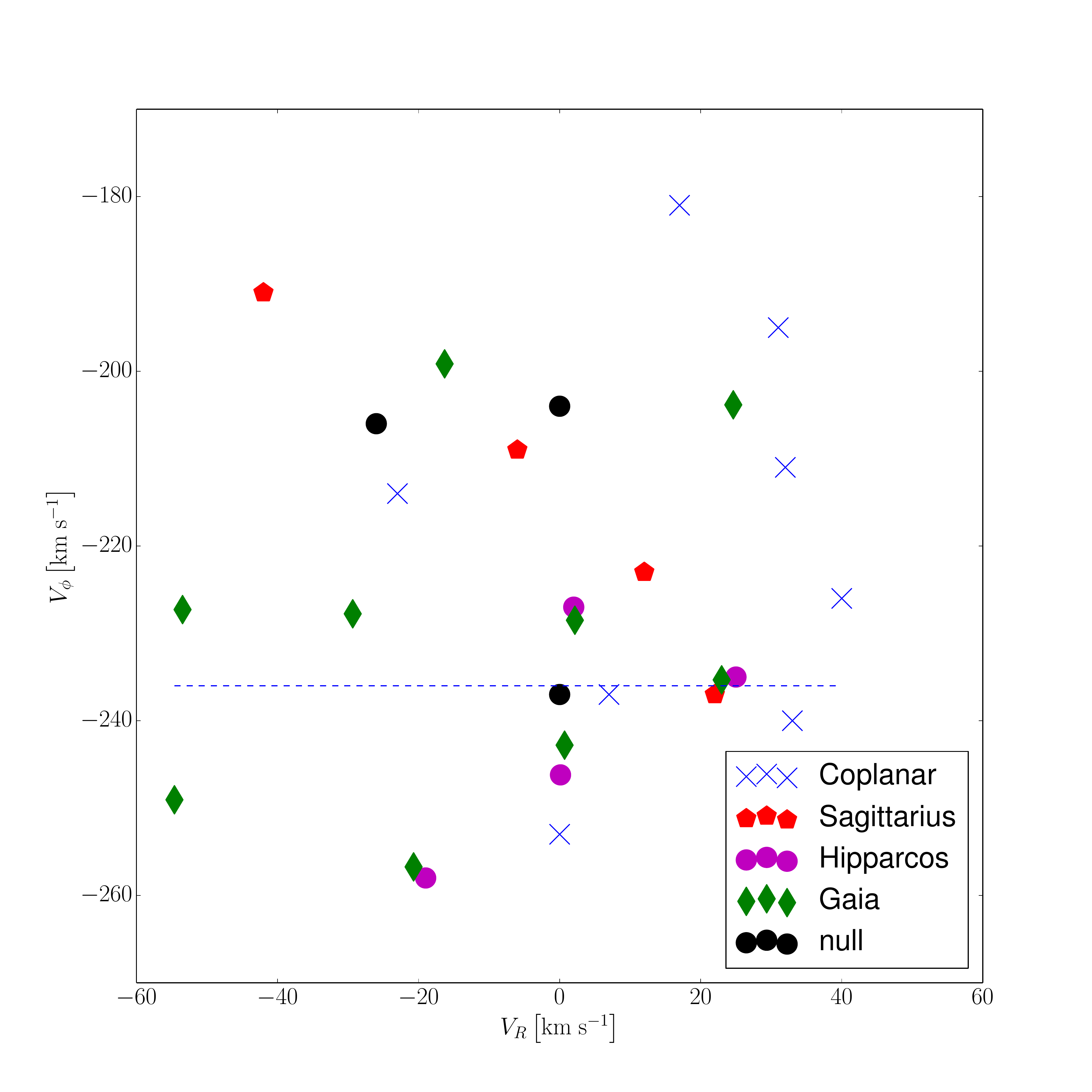}\\
\caption{This scatter plot shows the locations of moving groups in Gaia data,  the Hipparcos data, the Sagittarius simulation, the null simulation and in the co-planar simulation. Our simulations with dwarf galaxies match the dispersion in the Gaia data quite well in the $V_\phi$ direction, but the dispersion is somewhat smaller in the $V_R$ direction. The dispersion of the null simulation is smaller in both directions. The standard deviations of the group locations shown in this figure can be seen in Table \ref{spread}. The dashed horizontal line indicates the circular velocity.}\label{f:nullrad}
\end{center}
\end{figure}

\begin{table}
\centering
\begin{tabular}{c|c|c|c}
\hline\hline
Dataset &$ \sigma_U$ (km s$^{-1}$) & $\sigma_V$ (km s$^{-1}$) & Number of Groups\\
\hline
Gaia & 43.9 & 21.3 & 9\\
Sagittarius &  24.4 & 20.1 & 4\\
Co-planar & 20.0 & 22.7 & 8\\
Null &  12.2 & 15.1 & 3\\
Hipparcos & 15.6 & 11.7 & 4\\
\hline
\end{tabular}
\caption{This table contains the standard deviations of the locations of the groups displayed in Figure \ref{f:nullrad}}\label{spread}
\end{table}

To explore the possibility that there may be global aspects of dwarf galaxy interactions that leave imprints in velocity space, we have plotted the locations of moving groups detected by our algorithm in Gaia DR2 data in Figure \ref{f:nullrad}, and compared to the moving groups from our simulations at present day. The simulation groups displayed take data from four locations around the disk, to increase the size of the sample enough to provide a reasonable comparison to the Gaia data. The fiducial region displayed in previous figures is one of the regions selected, so that the groups shown there are also included here. The Gaia groups have a standard deviation in $V_R$ velocities that is about a factor of 1.5 larger than what is manifest in the co-planar simulation, while the standard deviation in $V_\phi$ for this simulation agrees with that in Gaia DR2 to better than 10 \%. The simulation with the three satellites has a standard deviation in $V_\phi$ that agrees with the Gaia data to better than a factor of 2, while the standard deviation in $V_R$ is within 25 \% for this case. In contrast, the null simulation in both $V_R$ and $V_\phi$ is discrepant with the Gaia data, at the level of a factor of 4 and a factor of 1.4, respectively.  The Hipparcos data shows smaller standard deviations, which is caused by the smaller number of groups detected in this sample. The standard deviations for all of the samples seen in Figure \ref{f:nullrad} can be found in Table \ref{spread}. The Gaia data can cover a similar spatial volume available as the portion of the simulations used, and in fact extends beyond the 1 kpc limit placed on the simulated data. The sample used for the Gaia sample here is not volume limited, so we have used all of the available sources. While the simulations that we have analyzed here do not span a large enough parameter space for us to draw definitive conclusions, the comparison of the dispersion in $V_R-V_\phi$ velocities in the Gaia DR2 data with the interacting simulations raises the question whether it may be possible to characterize dwarf galaxy interactions (to some extent) by analyzing the velocity spread in the moving group structures formed by such interactions.  Nevertheless, the discrepancy between the null simulation and Gaia DR2 data, and closer agreement between the interacting simulations and the moving groups in Gaia DR2 data shows that the standard deviation in 
$V_{r}$, $V_{\phi}$ space is better reproduced by dynamical interactions with dwarf galaxies than the isolated simulations that we have studied here. This is also true for the Hipparcos sample, which is a better match to the interacting simulations than to the null simulations. This result is expected as the largest moving groups should have similar properties in the more local sample from Hipparcos. The data set from Hipparcos covers a significantly smaller volume than the Gaia case or the samples from the simulations. However, since here we are measuring the deviation of the central group locations, the volume of the sample is not expected to impact the results.

\section{Discussion \& Conclusion}

In this paper, we compared moving groups identified in a series of simulations with data from the Milky Way, using an algorithm that we developed to identify larger groups in the $U-V$ plane.  We compared the results of our algorithm to Hipparcos data and to Gaia DR2 data, and we find that our algorithm successfully detects and recovers the parameters of a number of large moving groups.  However, it is unable to detect the smaller moving groups. By analyzing our simulations, we find that without a dwarf galaxy interaction, there is significantly less group formation than in simulation that includes gravitational interactions with dwarf galaxies, which suggests that some of the groups in the Milky Way could form from external perturbations. Yet another aspect of the dynamical formation nature of the moving groups is that the groups begin to appear in the disk within a dynamical time after the collision.  Without a dwarf galaxy, the algorithm still finds occasional moving groups, but finds an average of 0.1 structures per data set, which is lower than what we find in the interacting simulations.

Our simulations do not generate as many moving groups as we observe near the Sun in the Milky Way.
There are several explanations for this.  It is likely that this difference is due to the number of perturbing mechanisms, both external and internal to the Galaxy. A larger number of perturbation sources (i.e., in addition to dwarf galaxies, such as globular clusters or molecular clouds or bars) may very well be able to produce a larger number of moving groups, and thus produce something closer to that observed in the Milky Way.  None of the simulations form a bar, so none of the moving groups in the simulations are formed from a bar interaction. This may explain why we see fewer groups in the simulations than we do in the Gaia data. Alternatively, the larger number of Milky Way moving groups could result from a larger number of dwarf galaxy satellites. Moving groups can also be formed as the remnants of open clusters.  We can distinguish between groups that are open cluster remnants and dynamically formed groups by comparing the ages of the stars in the groups. If the stars have a wide range of ages, then it most likely formed through some kind of dynamic process, such as a dwarf galaxy perturbation. It is also likely that we would be able to find more moving groups in the simulations by increasing the resolution.  The groups themselves are in general quite similar between observations and simulations, i.e., the fractions and standard deviations of the groups are typically within a factor of two relative to the observed values.

Another feature that is worth mentioning is the dispersion of the moving groups in the $V_{r}, V_{\phi}$ plane. In the Hipparcos data, the moving groups are centered relatively close to the LSR. In our simulations, this is not the case, in fact there are groups that extend tens of kilometers per second beyond the farthest groups found within Hipparcos, particularly in the $V$ direction. Using the Gaia DR2 data, this discrepancy reverses, with the observed data having a higher standard deviation in the $V$ direction, which is a better match to the results in our simulations, particularly those of the co-planar interaction modeled by \cite{Chakrabarti&Blitz}, which has parameters similar to the Antlia 2 interaction \citep{2019arXiv190604203C}.  If this dispersion in $V$ velocities is particular to the co-planar interaction, and is not characteristic of other interactions (as the Sgr dwarf interaction), it suggests that velocity tracers of the predicted dwarf galaxy from \cite{Chakrabarti&Blitz} could be present in the Gaia DR2 data set but not in the older Hipparcos data set, due to the smaller volume of the Hipparcos sample.  

A significant part of this work has been the development of an algorithm that is useful for identifying large moving groups. This algorithm has identified a number of structures within the results of our simulations. In a typical snapshot for the interacting simulations, we can identify several large structures looking at a Galactocentric radius similar to that of the sun. These groups will typically have standard deviations on the order of a few $km s^{-1}$, and contain from 2 - 10 $\%$ of the total number of particles in the sample.

There are a few potential sources of future work here. One is to study the spatial structure of the groups found in the simulations. This may be easier with higher resolution, and could provide some additional insight into the origins of moving groups in the MW. Similar methods could also be applied to a set of simulations with higher resolutions and potentially different satellite parameters, which could also provide interesting additional information.

\section*{Acknowledgements}

Part of this work was conducted at the KITP Santa Barbara long-term program ``Dynamical Models for Stars and Gas in Galaxies in the Gaia Era".  SC and HN gratefully acknowledge the hospitality of the KITP during their visit, and NSF PHY-1748958.  The simulations have been performed on Xsede, NERSC, and on Google Cloud.  SC gratefully acknowledges support from NASA ATP NNX17AK90G, NSF AAG grant 1517488, and from Research Corporation for Scientific Advancement's Time Domain Astrophysics Scialog. HJN acknowledges 
support from NSF grants AST 16-15688 and AST 19-08653. This work has made use of data from the European Space Agency (ESA)
mission {\it Gaia} (\url{https://www.cosmos.esa.int/gaia}), processed by
the {\it Gaia} Data Processing and Analysis Consortium (DPAC,
\url{https://www.cosmos.esa.int/web/gaia/dpac/consortium}). Funding
for the DPAC has been provided by national institutions, in particular
the institutions participating in the {\it Gaia} Multilateral Agreement.

\section*{Data Availability}
The data underlying this article will be shared on reasonable request to the corresponding author.

\bibliographystyle{mnras}
\bibliography{bibl}

\bsp	
\label{lastpage}
\end{document}